\documentclass[prb,superscriptaddress,floatfix,a4paper,aps,twocolumn,nofootinbib]{revtex4-1}

\usepackage{graphicx}
\usepackage{amsmath}
\usepackage{amssymb}
\usepackage[usenames, dvipsnames]{xcolor}
\usepackage[colorlinks=true,citecolor=blue,linkcolor=blue,urlcolor=blue]{hyperref}

\usepackage{tabularx}
\usepackage{color}
\usepackage{xspace}

\begin{document}

\title{Huge linear magnetoresistance due to open orbits in $\gamma$-PtBi$_2$}

\author{Beilun Wu}
\affiliation{Laboratorio de Bajas Temperaturas y Altos Campos Magn\'eticos, Departamento de F\'isica de la Materia Condensada, Instituto Nicol\'as Cabrera and Condensed Matter Physics Center (IFIMAC), Universidad Aut\'onoma de Madrid, E-28049 Madrid,
Spain}

\author{V\'ictor Barrena}
\affiliation{Laboratorio de Bajas Temperaturas y Altos Campos Magn\'eticos, Departamento de F\'isica de la Materia Condensada, Instituto Nicol\'as Cabrera and Condensed Matter Physics Center (IFIMAC), Universidad Aut\'onoma de Madrid, E-28049 Madrid,
Spain}

\author{Hermann Suderow}
\affiliation{Laboratorio de Bajas Temperaturas y Altos Campos Magn\'eticos, Departamento de F\'isica de la Materia Condensada, Instituto Nicol\'as Cabrera and Condensed Matter Physics Center (IFIMAC), Universidad Aut\'onoma de Madrid, E-28049 Madrid,
Spain}
\affiliation{Unidad Asociada de Bajas Temperaturas y Altos Campos Magn\'eticos, UAM, CSIC, Cantoblanco, E-28049 Madrid, Spain}

\author{Isabel Guillam\'on*}
\affiliation{Laboratorio de Bajas Temperaturas y Altos Campos Magn\'eticos, Departamento de F\'isica de la Materia Condensada, Instituto Nicol\'as Cabrera and Condensed Matter Physics Center (IFIMAC), Universidad Aut\'onoma de Madrid, E-28049 Madrid,
Spain}\affiliation{Unidad Asociada de Bajas Temperaturas y Altos Campos Magn\'eticos, UAM, CSIC, Cantoblanco, E-28049 Madrid, Spain}

\begin{abstract}
Some single-crystalline materials present an electrical resistivity which decreases between room temperature and low temperatures at zero magnetic field as in a good metal and switches to a nearly semiconductinglike behavior at low temperatures with the application of a magnetic field. Often, this is accompanied by a huge and nonsaturating linear magnetoresistance which remains difficult to explain. Here, we present a systematic study of the magnetoresistance in single-crystal $\gamma$-PtBi$_2$. We observe that the angle between the magnetic field and the crystalline $c$ axis fundamentally changes the magnetoresistance, going from a saturating to a non-saturating magnetic-field dependence. In between, there is one specific angle where the magnetoresistance is perfectly linear with the magnetic field. We show that the linear dependence of the nonsaturating magnetoresistance is due to the formation of open orbits in the Fermi surface of $\gamma$-PtBi$_2$. 

\end{abstract}

\maketitle

%%%%%%%%%%%%%%%%%%%%%%%%%%%%%%%

%%%%%%%%%%%%%%%%%%%%%%%%%%%%%%%

Magnetoresistance (MR) is the modification of the electrical resistance by a magnetic field. MR is a ubiquitous phenomenon in metals and semiconductors, although it is not expected to occur just considering free electrons without interactions. The electrical resistivity $\rho$ occurs due to scattering of electrons on a timescale $\tau$, and the main consequence of applying a magnetic field is spiraling the electron orbits with an angular velocity $\omega_c=\frac{eB}{m*}$ (with $e$ as the electron charge and $m*$ as the electronic effective mass). When considering nearly free electrons, the simplest magnetic field dependence  found for the magnetoresistance is quadratic $\rho(B) \propto B^2$, obeying Onsager's reciprocity condition $\rho(B)=\rho (-B)$.  Furthermore, the MR saturates in the high field limit ($\omega_c \tau \gg 1$) unless electron and hole numbers are close to compensate with each other, in which case it continues growing as $\rho(B)\propto B^2$. In metals with no or weak electronic correlations, a small quadratic MR saturating at high fields is indeed often observed\cite{PippardBook,AbrikosovBook,Kapitza}, with a few exceptions such as the semimetal Bi and other electron-hole compensated metals\cite{Zhu_2018,PhysRev.91.1060,PhysRevLett.82.3328,Ali2014}. More intriguing and difficult to explain is the observation of a huge and non-saturating linear magnetoresistance, which has been recently discussed in a number of materials. The magnetoresistance is also influenced by contributions from open orbits, which play sometimes a role, often leading to a sizable enhancement, which was however not considered as sufficient to explain huge magnetoresistances\cite{PippardBook,AbrikosovBook,Kapitza}. Here, we show that the direction of the magnetic field produces an unanticipated huge enhancement of the linear magnetoresistance exactly when the contribution of open orbits to the magnetoresistance is maximal.

%%%%%%%%%%      Figure 1     %%%%%%%%%%%%

\begin{figure*}[htbp]
\includegraphics[width = 2\columnwidth]{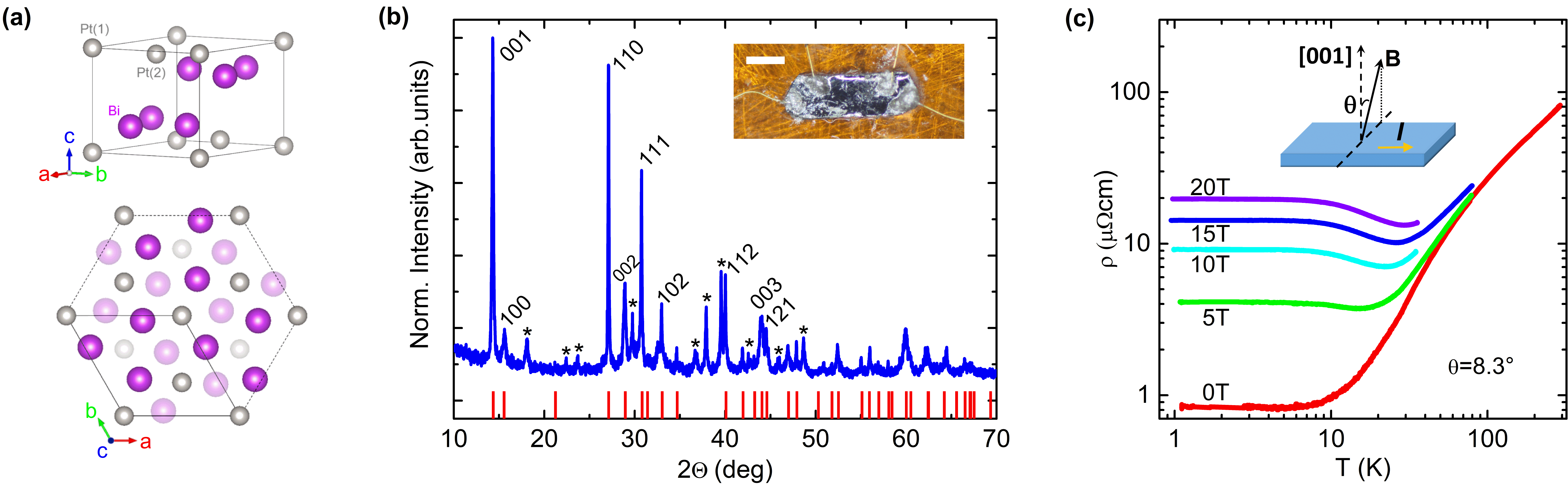}
\caption{
\textbf{(a)} Crystal structure of the layered $\gamma$-PtBi$_2$. 
\textbf{(b)} In blue we show the x-ray diffraction pattern of $\gamma$-PtBi$_2$ powder. Red bars show the positions of the peaks expected to appear in this compound. The asterisks mark the peaks associated with residual Bi and Bi oxides from flux growth. The inset shows a picture of the single crystal with four contacts used for resistivity measurements. The white scale bar is 0.2 mm long. 
\textbf{(c)} Colored lines show the temperature dependence of the resistivity at different magnetic fields. The field is applied at an angle $\theta = 8.3^{\circ}$, which is also the precise angle at which we find nonsaturating linear magnetoresistance. The temperature dependence is very similar for all field orientations. 
%There is a strong decrease in the resistivity with decreasing temperature at zero magnetic field, which turns into an increase when applying the magnetic field. 
The inset shows a scheme of the direction of the applied current and magnetic field.
}
\label{FigureXRay}
\end{figure*}

There are a number of metallic or semimetallic compounds showing large and sometimes linear magnetoresistances\cite{Tafti2015,Feng11201,Leahy10570,PhysRevB.99.045119,PhysRevB.57.13624,MYERS199927,Ali2014,PhysRevB.85.035135,Wu2016,PhysRevB.92.235153,PhysRevB.87.214504,PhysRevB.96.235128,Shekhar2015,Lv2017,Ma2018,PhysRevB.96.165145}. The semimetal $\gamma$-PtBi$_2$ stands out among these compounds because of the extreme values of the magnetoresistance\cite{Gao2018}. $\gamma$-PtBi$_2$ has a layered structure with trigonal symmetry (space group $P$31$m$, No.157, see Fig.\,\ref{FigureXRay}(a)). Electronic band structure calculations\cite{Xu2016, Gao2018} show that this compound has a Fermi surface containing multiple electron and hole sheets. Angle-resolved photoemission spectroscopy (ARPES)\cite{Yao2016,Thirupathaiah2018} and quantum oscillation studies\cite{Gao2018} have measured the band structure and the Fermi surface. In particular, the ARPES data\cite{Thirupathaiah2018} revealed a spin-polarized surface state with linear dispersion, which was associated with the linear MR reported in an early study\cite{Yang2016}. 

Here, we make detailed measurements of the angular dependent MR up to 22 T on a high quality single crystal of $\gamma$-PtBi$_2$. 
%We apply the current in-plane ($\perp \mathbf{c}$-axis) and rotate the magnetic field away from the $\mathbf{c}$-axis into the plane. % repeated in the methods.
To analyze effects only due to the crystalline orientation to the magnetic field, we keep the field direction perpendicular to the electrical current. We observe a continuous evolution from a saturating sublinear MR for $\mathbf{B}\parallel \mathbf{c}$ to a nonsaturating quadraticlike MR for the field on the plane. The linear nonsaturating MR is only observed for a specific angle of the magnetic field with respect to the $c$ axis. We show that such a linear MR appears at a specific angle in the presence of open orbits.

We grew single crystals of $\gamma$-PtBi$_2$\cite{osti_7157035} using the Bi self-flux method described in Ref.\onlinecite{Gao2018}. We used the equipment described in Ref.\onlinecite{Canfield1992}, in particular, frit crucibles\cite{Canfield2016}. Powder x-ray diffraction of our crystals reveals the expected crystal structure (Fig.\ref{FigureXRay}(b)), together with some peaks corresponding to residual Bi flux and Bi oxides. We measured a neat $\gamma$-PtBi$_2$ single crystal platelet, oriented with the $\mathbf{c}$-axis out of the plane (inset of Fig.\,\ref{FigureXRay}(b)). The residual resistance ratio is of 100, showing excellent sample quality. To measure the MR, we used a cryostat capable of reaching about 1 K\cite{MONTOYA2019e00058}, and a 20+2 T superconducting magnet supplied by Oxford Instruments\cite{Magnet}. We used a homemade mechanical rotator, described in the Supplemental Material\cite{Supp}, to modify the field angle. The current was applied perpendicular to the magnetic field (inset in Fig.\ref{FigureXRay}(c)). The rotator allowed an angular range covering from the field along the $c$ axis ($\theta=0^\circ$) to parallel to the plane ($\theta=90^\circ$). The angle of the magnetic field was measured using a Hall probe\cite{HallProbe}. We define MR$=\frac{\rho(B)-\rho(0)}{\rho(0)}$ with $\rho(0)$ being the resistivity at zero magnetic field, and provide it in percentage. To find Shubnikov de Haas oscillations, we obtain the oscillatory component $\Delta$MR by fitting the MR data above 13 T to a low order polynomial and perform the Fourier transform of $\Delta$MR$(1/B)$.

Fig.\,\ref{FigureXRay}(c) shows the resistivity as a function of temperature at different magnetic fields. We find a metallic behavior at zero field 
%(resistivity strongly decreases with decreasing temperature) 
and a semiconductinglike behavior under magnetic fields (resistivity increases with decreasing temperature at low temperatures). The semiconducting-like increase saturates below about 8 K.

%%%%%%%%%%   Figure 2   %%%%%%%%%%%

\begin{figure}[h!]
\includegraphics[width = \columnwidth]{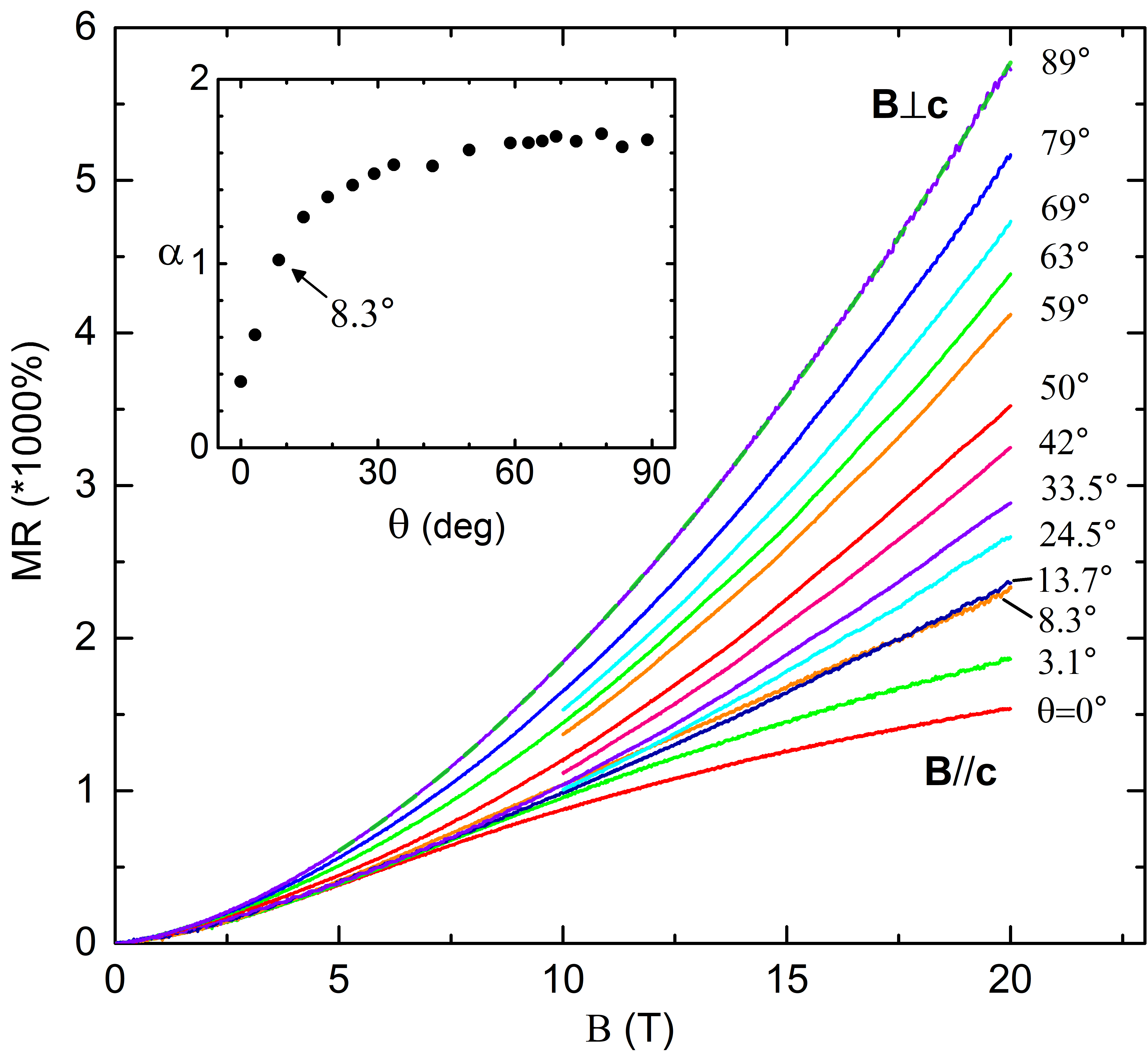}
\caption{MR up to 20 T for different field directions from $\mathbf{B}\parallel \mathbf{c}$ ($\theta = 0^{\circ}$) to $\mathbf{B}$ $\perp \mathbf{c}$ ($\theta$ close to $90^{\circ}$). The inset shows the angular dependence of the exponent $\alpha$ obtained by fitting the MR at each angle with an empirical power law MR\,$= c +a B^{\alpha}$. The green broken line in the main figure shows, as an example, the fit for the data at $\theta=89^{\circ}$.}
\label{FigureMR}
\end{figure}

%%%%%%%%%%   Figure 3   %%%%%%%%%%%

\begin{figure*}[htbp]
\includegraphics[width = 2\columnwidth]{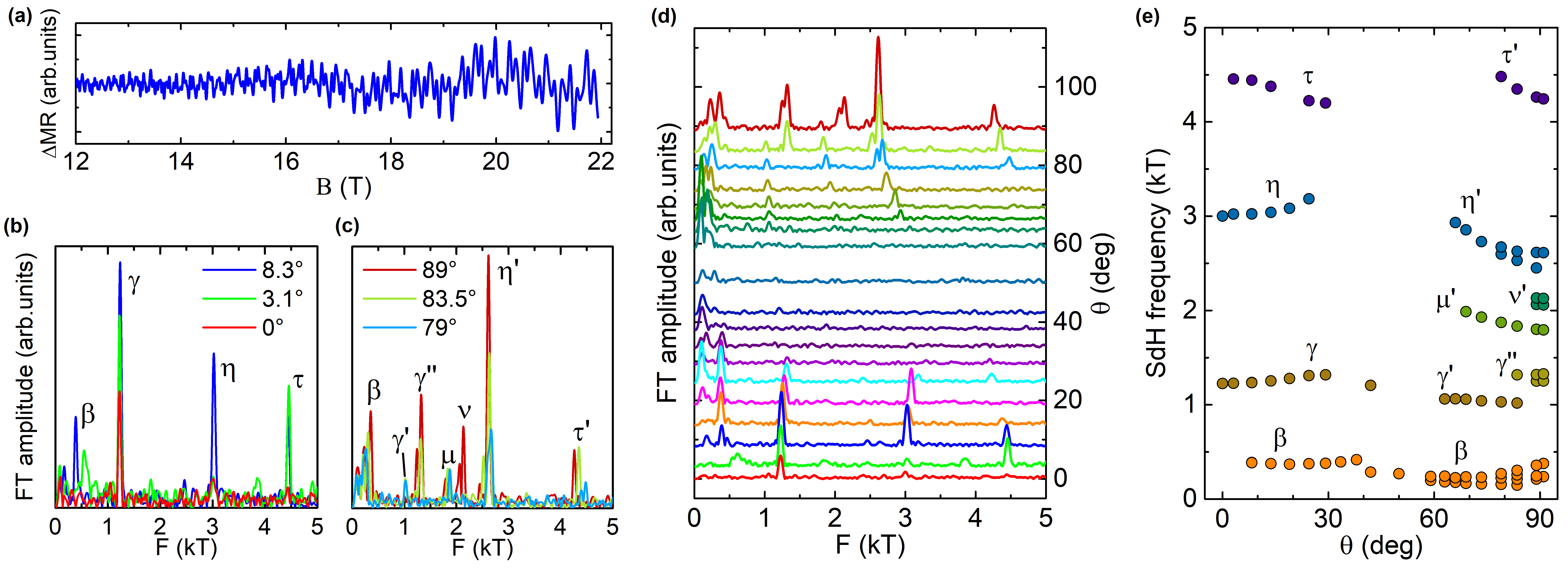}
\caption{\textbf{(a)} $\Delta $MR (defined in the text) versus magnetic field up to 22 T at T=1 K and $\theta = 8.3^{\circ}$.
\textbf{(b,c)} Fourier transform of $\Delta$MR for a few values of $\theta$. Greek letters mark the oscillation frequencies giving peaks in the Fourier transforms. \textbf{(d)} Fourier transform of $\Delta$MR as a function of the frequency $F$ for different $\theta$. Data are shifted along the $y$-axis for clarity, following the value of $\theta$'s. \textbf{(e)} Angular evolution of the quantum oscillations frequencies, with the corresponding orbits labeled by greek letters.}
\label{FigOsc}
\end{figure*}

Fig.\,\ref{FigureMR} shows the transverse MR up to 20 T for different orientations of the magnetic field, from $\mathbf{B}\parallel \mathbf{c}$ ($\theta = 0^{\circ}$) to $\mathbf{B}$ $\perp \mathbf{c}$ ($\theta = 90^{\circ}$). The highest value of the MR is of $5800\%$ at 20 T for $\theta$ close to $ 90^{\circ}$. For magnetic fields below 5 T, we always find a quadratic MR. For higher fields, we observe an angular evolution of the MR, from a concave curvature (saturating MR) at $\theta = 0^{\circ}$ to a quadratic-like, convex curvature (nonsaturating MR) at $\theta$ close to $90^{\circ}$. The magnitude of the MR changes by a factor of 5 as a function of the angle at high magnetic fields. In between the concave and convex magnetic field dependencies we observe a perfectly linear MR at $\theta=8.3^\circ$.

To see this, we fit the data above 5 T with an empirical power law MR$(B)= c+aB^{\alpha}$. We find an increasing $\alpha$ with $\theta$ (inset of Fig.\,\ref{FigureMR}) up to $1.65$. At $\theta=8.3^\circ$, $\alpha=1$ and the MR changes from saturating to non-saturating behavior. 

Fig.\,\ref{FigOsc}(a) shows the quantum oscillation pattern in the MR for $\theta=8.3^{\circ}$ as a function of the magnetic field. Fig.\,\ref{FigOsc}(b-d) present the Fourier transform of the quantum oscillation signal as a function of $\theta$.  Each quantum oscillation frequency $F$ is related to an extremal cross-sectional area of the Fermi surface normal to the field $\mathcal{A}_k$ through the Onsager relation, $F= (\hbar/2 \pi e)\mathcal{A}_k$. In Fig.\,\ref{FigOsc}(e), we track the quantum oscillation frequencies as a function of the angle. At $\theta$ close to $0^\circ$ our result exactly coincides with the result in Ref.\onlinecite{Gao2018} (taken along the same field direction). We can thus identify several frequencies, F$_{\beta}$=388 T, F$_\gamma$=1225 T and F$_\eta$=3012 T, using the same notation\cite{Gao2018}.  The results for finite $\theta$ are new and we use the same notation extrapolating from $\theta=0^\circ$. We measure in a smaller field range than Ref.\onlinecite{Gao2018}. As a result two low frequency orbits (at 40 and at 15 T) are not well defined in our data. On the other hand we can sweep the magnetic field much more slowly and thus resolve better the high frequency quantum oscillations. For instance, we observe a high frequency oscillation around  F$_\tau$=4440 T. The quantum oscillation pattern is very weak between $\theta=30^{\circ}$ and $60^{\circ}$, so we cannot resolve the frequency of the orbits in this angular range, except for the orbit with the lowest frequency, $\beta$. We, thus, keep the notation for the orbits at $\theta$ close to $90^\circ$ as for similar frequencies at $\theta$ close to $0^\circ$ but add apostrophes. There are two extra frequencies that do not have direct counterparts at $\theta$ close to $0^\circ$, and we denote them as $F_{\mu'}$=1798 T and $F_{\nu'}$=2012/2134 T. Furthermore, we observe that $\eta'$, $\gamma''$ and $\nu'$ split into two and $\beta$ into three frequencies close to $90^\circ$.

We have also measured the temperature dependence of the quantum oscillations at two different angles $\theta = 8.3^\circ$ and $\theta = 89^\circ$ where the oscillation amplitudes are the largest, to obtain the quasiparticle effective mass $m_i^*$ of each orbit and their quantum lifetime $\tau_Q$ (see Fig.\,S1(a)-S1(c) and Table S1 in the Supplemental Material\cite{Supp}). $m_i^*$ lies close to the free-electron mass $m_e$ for all the orbits except $\beta$, whereas the estimated $\tau_Q$ ranges from 0.15 to 0.75 ps. We also analyze the phase of the lowest-frequency mode ($\beta$) and obtain, for this orbit, a nontrivial Berry phase $\Phi_B$ close to $\pi$. This result is in agreement with previous quantum oscillation measurements and suggests that the predicted triply degenerate point in the associated $\beta$ band is likely to be present\cite{Gao2018}.

%Our results show a continuous change of the MR curvature, from saturating to non-saturating, when the magnetic field direction is rotated with respect to the crystalline axes. The data reveal a perfectly linear, non-saturating MR at one specific field angle. Furthermore, we have measured new quantum oscillation frequencies, corresponding to larger Fermi surface sheets, not reported in previous work, and found topologically non-trivial properties in part of the band structure.

Band structure calculations in $\gamma$-PtBi$_2$\cite{PhysRevB.94.165201,Gao2018} show many bands that cross the Fermi level, forming six or seven sheets. Spin orbit coupling leads to large band splittings and the shape of the Fermi surface pockets is very sensitive to the position of the Fermi level. At the Fermi level, the bands have mixed electron-hole character with a somewhat larger density of holes\cite{Yao2016,Thirupathaiah2018}. We see that the smallest $k_F$ ($\beta$ orbit) in our quantum oscillation data is about 1/4 of the size of the first Brillouin zone and appears in pockets centered at the corners of the Brillouin zone derived from the $\beta$ band\cite{Gao2018}. The two high-frequency orbits ($\eta$ and $\tau$) are associated with the two bands producing large pancakelike Fermi-surface sheets centered at the top and the bottom of the Brillouin zone\cite{Gao2018}. This particular shape gives extremal orbits only when $\mathbf{B}\parallel \mathbf{c}$ or $\perp\mathbf{c}$, which explains the absence of corresponding frequencies at intermediate angles. We also find several orbits split into two or more frequencies at $\theta$ close to $90^{\circ}$. This probably comes from the warping of the corresponding Fermi-surface sheets\cite{Carrington2005}.

%%%%%%%%%%   Figure 4   %%%%%%%%%%%

\begin{figure}[h!]
{\includegraphics[width =\columnwidth]{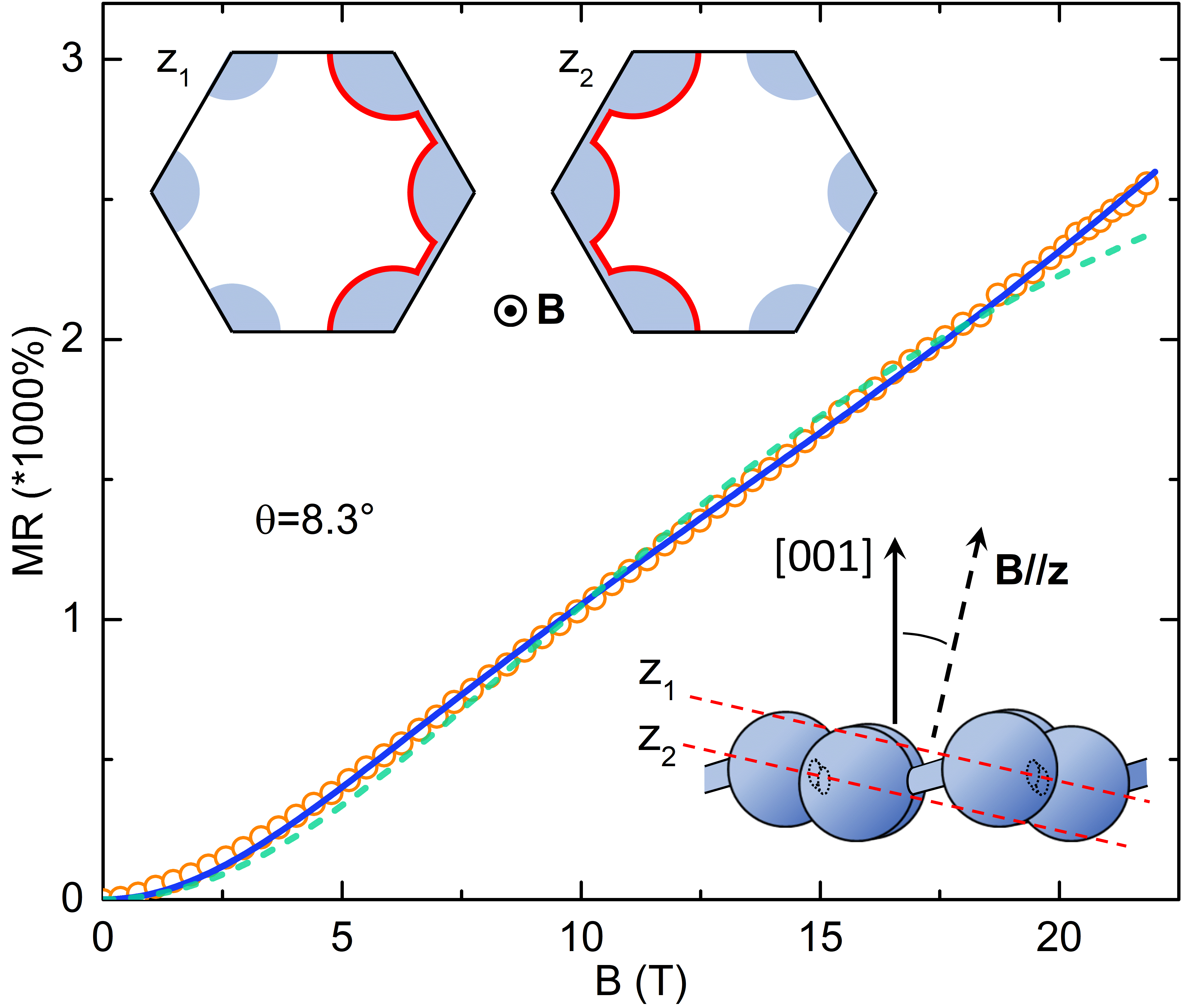}}
\caption{We show as orange circles the MR at $\theta = 8.3^{\circ}$ up to 22 T. By the blue line we show the result of the model discussed in the text. The green dashed line shows the result of the same model, but without contribution from open orbits. Note the saturation observed at high magnetic fields. In the lower right inset, we show schematically the $\gamma$-band Fermi surface sheet. The shape is that of an ondulated grid of spheres arranged in a honeycomb lattice perpendicular to the $\mathbf{c}$ axis\cite{Gao2018}. The sheets are connected by necks oriented at an angle to the plane. Necks that lie behind the spheres are schematically represented by dashed lines. When the magnetic-field direction is slightly tilted from $\mathbf{c}$, open orbits may appear on two different planes (marked by red dashed lines). The corresponding open orbits are provided in the upper left panels.
}
\label{Explanation}
\end{figure}

The calculations in Ref.\onlinecite{Gao2018} reveal a $\gamma$ band that has sphericallike Fermi-surface pockets arranged in a honeycomb lattice located on the plane and interconnected with each other through tilted necks. The lower inset of Fig.\,\ref{Explanation} presents schematically this open Fermi surface sheet. We mark, by the red dashed lines, two planes perpendicular to a magnetic field tilted from the $c$ axis. These contain open orbits. In each of the planes, the spheres located at one side of the honeycomb structure are connected to each other, but not with the spheres of the other side (upper left insets of Fig.\,\ref{Explanation}). Electrons on open orbits go into non-circular trajectories, instead of following cyclotron motion\cite{Lifshitz1957,PippardBook}. This has a strong impact on the MR\cite{Lifshitz1957,Lifshitz1959,Lifshitz1960,PippardBook}. In a metal with a single electronic band, open orbits lead to a quadratic enhancement of the MR at certain field angles, leaving a usual saturated MR at the angles where the open orbits are absent\cite{Klauder1966,Huberman1982}. Multiple Fermi-surface sheets as in $\gamma$-PtBi$_2$ are considered in detail here. To set up a MR model taking into account multiple sheets, we first consider that, in a semimetal, the MR has a $B^2$ unsaturated behavior when the electron and hole numbers ($n_e$ and $n_h$) are compensated\cite{PippardBook,Ali2014,Gao2017}. We consider a two-band model and take electrons as free carriers with the same mobility $\mu$. The level of electron-hole compensation is given by $d=(n_e-n_h)/(n_e+n_h)$. We add a field-independent small contribution from the open orbits $\delta \sigma _0$ to the total conductivity $\sigma _0$. More details are provided in the Supplemental Material\cite{Supp}. We can then write for the resistivity $
\rho(B)/\rho_0=\frac{\delta(1+\eta^2)^2+(1+\eta^2)}{\delta (1+\eta^2)+1+ d^2\eta^2}$, where $\eta=\mu B$.

We find that this reproduces nicely the linear MR at $8.3^\circ$ (Fig.\,\ref{Explanation}) and at all other field orientations (see Fig.\,S2 in the Supplemental material\cite{Supp}). We use for this angle the parameters $\delta = 0.00849$, $\mu = 4630$ cm$^2$V$^{-1}$s$^{-1}$) and $|d| = 0.229$. The size of $\delta$ is small compared to one, in agreement with the fact that only a tiny proportion of the whole Fermi surface would be engaged in forming open orbits. 
%This means that only a tiny proportion of the Fermi surface is engaged in forming the open orbits, in agreement with the previous discussion (and the fact that the maximum in the quantum oscillation amplitude occurs for $\theta=8.3^\circ$ due to the open orbits contribution to the MR). 
From the mobility $\mu$, we estimate the transport mean free scattering time $\tau_{tr}$ ($\mu = e\tau_{tr}/m^*$) to be around $2.6\cdot 10^{-12}$s, a few times larger than the quantum lifetime $\tau_Q$ deduced from the amplitude of the quantum oscillations (Table S1 in the Supplemental Material\cite{Supp}). This difference is reasonable as $\tau_Q$ measures the smearing of the Landau levels due to forward and backward scatterings, whereas transport $\tau_{tr}$ requires backscattering. The linear MR cannot be obtained without taking into account the open orbits ($\delta$ term), and the MR would show a clear saturation at high fields due to an inexact compensation between the electron and the hole carrier numbers. It is, thus, the combination of both effects that leads to the observed linear MR at $\theta=8.3^{\circ}$. For the other angles, we can associate the increase in the MR from $0^\circ$ to $90^\circ$ and the evolution of its curvature to changes in $|d|$, which steadily decreases from 0.22 at  $0^\circ$ to close to 0 at $90^\circ$. 

The relevant contribution of a one-dimensional conduction channel for the magnetoresistance which we unveil here can have consequences, particularly, for the behavior at even larger magnetic fields. The Lebed effect consists of resonances occurring in open orbits for in- and out-of-plane transport between layers of a two-dimensional crystalline structure \cite{PhysRevLett.63.1315}. As a consequence, different kinds of oscillatory behavior related to open orbits have been predicted and observed in organic systems, including magic angles related to dimensional crossovers, strong angular-dependent oscillations, or density waves\cite{PhysRevLett.105.067201,PhysRevLett.93.157006,PhysRevLett.72.3714,doi:10.1143/JPSJ.75.051006}. Our result suggests that such one-dimensional features could arise in semimetals due to open orbits.

This seems particularly interesting in view of the unconventional bandstructures often observed in semimetals. For $\gamma$-PtBi$_2$, band structure calculations suggest triply degenerate nodal points \cite{Gao2018,Lv2017,Ma2018,Thirupathaiah2018}. Contrary to Dirac and Weyl fermions\cite{Bradlynaaf5037,PhysRevB.94.165201,Gao2018,Chang2017}, triple points have no counterpart in high energy physics\cite{Bradlynaaf5037,Chang2017,PhysRevB.94.165201,PhysRevB.99.245118}. These could influence the magnetoresistance in the case of complex conduction paths involving open orbits induced by the magnetic field.

To summarize, we synthesized high quality single crystals of $\gamma$-PtBi$_2$ and measured the angular dependence of the MR up to 22 T. We reveal that, in addition to the known electron-hole compensation, open orbits produce an important modification of the transport under magnetic fields leading to huge values of the magnetoresistance and a linear field dependence.
%The observed phenomenon is a compensation effect between contributions to the magnetoresistance of two kinds of carriers and the open orbits and adds to the known electron-hole compensation leading to the $B^2$ magnetoresistance as an effect providing huge magnetoresistances. 
Additionally, we suggest that phenomena inherent to one-dimensional conduction might appear at ultra high magnetic fields in low carrier density semimetals.

\section*{Acknowledgements}

This work was supported by the European Research Council PNICTEYES Grant Agreement No. 679080, by the Spanish Research State Agency (Grants No. FIS2017-84330-R, No. CEX2018-000805-M, and No. RYC-2014-15093) and by the Comunidad de Madrid through Program NANOMAGCOST-CM (Program No. S2018/NMT-4321). We acknowledge collaborations through the EU program Cost CA16218 (Nanocohybri). We particularly acknowledge SEGAINVEX at UAM for design and construction of cryogenic equipment. We also thank R. \'Alvarez Montoya and J.M. Castilla for technical support and S. Delgado for support with the single crystal growth. We learned of the search for new materials with P. C. Canfield and are very grateful to him for this. We acknowledge SIDI at UAM for support in sample characterization.

\bibliographystyle{apsrev4-1-titles}
%\bibliography{bib}


%%%%%%%%%% Merge with supplemental materials %%%%%%%%%%
\widetext
\clearpage
%\onecolumngrid
\begin{center}
\textbf{\large Supplemental Materials for ``Huge linear magnetoresistance due to open orbits in $\gamma$-PtBi$_2$''}
\end{center}
%%%%%%%%%% Merge with supplemental materials %%%%%%%%%%
%%%%%%%%%% Prefix a "S" to all equations, figures, tables and reset the counter %%%%%%%%%%
\setcounter{equation}{0}
\setcounter{figure}{0}
\setcounter{table}{0}
\setcounter{page}{1}
\makeatletter
\renewcommand{\theequation}{S\arabic{equation}}
\renewcommand{\thefigure}{S\arabic{figure}}
\renewcommand{\thetable}{S\arabic{table}}
%\renewcommand{\bibnumfmt}[1]{[S#1]}
%\renewcommand{\citenumfont}[1]{S#1}
%%%%%%%%%% Prefix a "S" to all equations, figures, tables and reset the counter %%%%%%%%%%

\author{Beilun Wu}
\affiliation{Laboratorio de Bajas Temperaturas y Altos Campos Magn\'eticos, Departamento de F\'isica de la Materia Condensada, Instituto Nicol\'as Cabrera and Condensed Matter Physics Center (IFIMAC), Universidad Aut\'onoma de Madrid, E-28049 Madrid,
Spain}

\author{V\'ictor Barrena}
\affiliation{Laboratorio de Bajas Temperaturas y Altos Campos Magn\'eticos, Departamento de F\'isica de la Materia Condensada, Instituto Nicol\'as Cabrera and Condensed Matter Physics Center (IFIMAC), Universidad Aut\'onoma de Madrid, E-28049 Madrid,
Spain}

\author{Hermann Suderow}
\affiliation{Laboratorio de Bajas Temperaturas y Altos Campos Magn\'eticos, Departamento de F\'isica de la Materia Condensada, Instituto Nicol\'as Cabrera and Condensed Matter Physics Center (IFIMAC), Universidad Aut\'onoma de Madrid, E-28049 Madrid,
Spain}
\affiliation{Unidad Asociada de Bajas Temperaturas y Altos Campos Magn\'eticos, UAM, CSIC, Cantoblanco, E-28049 Madrid, Spain}

\author{Isabel Guillam\'on*}
\affiliation{Laboratorio de Bajas Temperaturas y Altos Campos Magn\'eticos, Departamento de F\'isica de la Materia Condensada, Instituto Nicol\'as Cabrera and Condensed Matter Physics Center (IFIMAC), Universidad Aut\'onoma de Madrid, E-28049 Madrid,
Spain}\affiliation{Unidad Asociada de Bajas Temperaturas y Altos Campos Magn\'eticos, UAM, CSIC, Cantoblanco, E-28049 Madrid, Spain}

\maketitle

\section{Temperature and field dependence of the Shubnikov-de Haas oscillations}

\begin{figure}[h!]
\includegraphics[width =0.8\columnwidth]{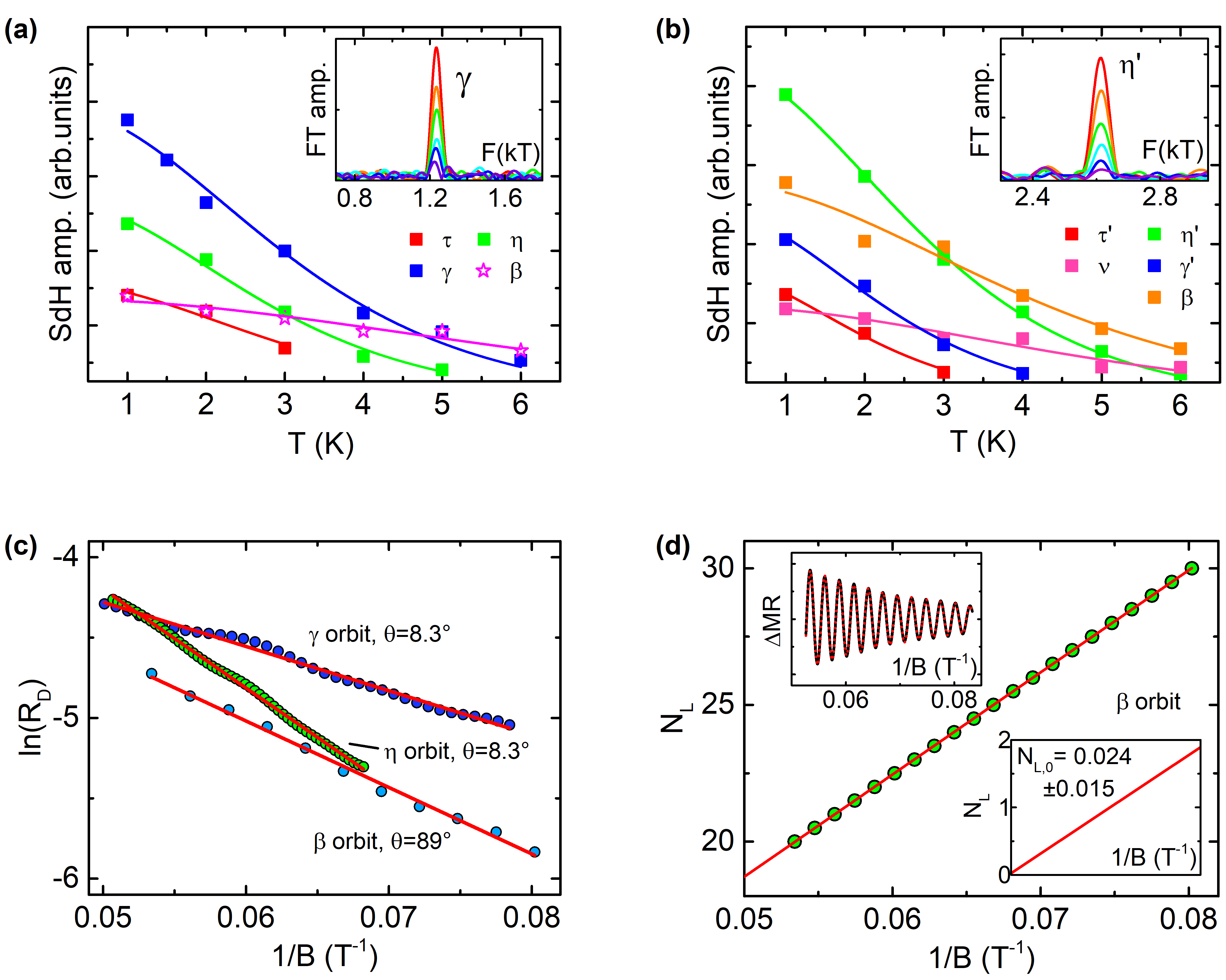}
\caption{
\textbf{(a,b)} We show as points the amplitude of the quantum oscillations of different frequencies observed at $\theta =8.3^{\circ}$ and $\theta =89^{\circ}$ as a function of temperature. Lifshitz-Kosevich fits are shown as solid lines. Each of the insets shows one of the Fourier transform peaks in the quantum oscillations with increasing temperature, from T=1 K (red) to 6 K (blue). \textbf{(c)} We show as points the Dingle factor $R_D$ (in a logarithmic scale) as a function of $1/B$ for a few selected frequencies ($\gamma$ at $\theta=8.3^{\circ}$, blue points, $\eta$ at the same angle, green points and $\beta$ at $\theta=89^{\circ}$, light blue points). The red lines are linear fits to each magnetic field dependence. \textbf{(d)} Landau level index ($N_L$) as a function of $1/B$ extracted from the peaks (integer) and valleys (half integer) for one of the $\beta$ orbits at $89^{\circ}$. The solid line is a linear fit. The lower right inset shows the zero $1/B$  intercept, giving $N_{L,0} = 0.024\pm 0.015$. The upper left inset shows the amplitude of quantum oscillations with frequency $F_{\beta}$ vs $1/B$ obtained by band-pass filtering the raw data (black points are the data and red dotted line the fit to LK formula) and used to obtain $N_L$.}
\label{Dingle}
\end{figure}

To discuss the temperature and magnetic field dependence of the amplitude of the quantum oscillations, let us write down the Lifshitz-Kosevich formula:

%%%%%%%%%%     Equation 1     %%%%%%%%%%

\begin{equation}
\label{eq_LK}
\Delta MR \propto \sqrt{B}\, \frac{\Theta m_i^* T/B}{\sinh (\Theta m_i^* T/B)}\,\exp{(-\Theta m_i^* T_D/B)}
\cos{\left[2\pi\left(\frac{F}{B}+\gamma-\delta\right)\right]}
\end{equation}

where $\Delta$MR is the oscillatory component of the MR, $m_i^*$ the effective electron mass of band $i$, $\Theta= 2\pi^2k_B/e\hbar$ is a numerical factor and $T_D=\hbar/2\pi k_B\tau_Q$ is the Dingle temperature. The phase factor $\cos{\left[2\pi\left(\frac{F}{B}+\gamma-\delta\right)\right]}$ depends on the frequency $F$ (normalized to $B$), on the phase shift $\delta$, given by the dimensionality of the Fermi surface (equal to 0 in a 2D system and $\pm$1/8 in 3D) and on the phase shift $\gamma$. $\gamma$ is related to the Berry phase $\Phi_B$ accumulated over the orbit as $\gamma=\frac{1}{2}-\frac{\Phi_B}{2\pi}$.

In Fig.\,\ref{Dingle}(a,b) we show the temperature dependence of the amplitude of the Fourier transforms of the quantum oscillations at $8.3^\circ$ and $89^\circ$, respectively. We obtain the quasiparticle effective mass $m_i^*$ of each orbit from the fits to equation \ref{eq_LK}. In Fig.\,\ref{Dingle}(c) we show the Dingle factor $R_D=\exp{(-\Theta m_i^* T_D/B)}$ as a function of $1/B$ for a few selected frequencies. From these fits, we obtain the quantum lifetime $\tau_Q$ in each band. The resulting parameters are summarized in Table\,\ref{TablePar}. The Fermi wavevector $k_F$ is calculated from the frequency $F$ with the Onsager relation and the assumption of a circular orbit: $\mathcal{A}=\pi k_F ^2$. We see that $k_F$ spans a large part of the Brillouin zone. The electron effective mass $m_i^*$ is very close to the free electron mass for nearly all orbits, except for the $\beta$ orbit where it is considerably smaller, about $0.4m_e$, in good agrement with previous quantum oscillation measurements\cite{Gao2018x}. The mean free path $\ell$ ranges from  400 \AA\ to about 4000 \AA. 

\begin{table}[h]
\caption{Fermi surface parameters obtained from the quantum oscillations for two particular angular orientations of the magnetic field. $F$ is the frequency, $k_F$ the Fermi wavevector, $m*$ the effective mass, $\ell(\AA)$ the quantum mean free path and $\tau _Q$ the quantum lifetime.}
\begin{tabular}{c ccc c    cc}

					&				&F(T)		&$k_F$(nm$^{-1}$)	&m$^*/m_e$				&$l$($\AA$)	&$\tau _Q$(ps)	\\
\hline
8.3$^{\circ}$		&$\beta$		&388		&1.09				& 0.39$\pm$0.03		&420			&0.14\\
					&$\gamma$	&1235		&1.94				&0.77$\pm$0.01		&1500			&0.51\\
					&$\eta$		&3023		&3.03				&0.90$\pm$0.03		&1020			&0.26\\
					&$\tau$		&4442		&3.67				&0.83$\pm$0.04		&3770			&0.73\\

89$^{\circ}$			&$\beta$		&377		&1.04				&0.61$\pm$0.03		&520			&0.26\\
					&$\gamma$'	&1323		&2.00				&1.00$\pm$0.01		&1070			&0.46\\
					&				&1248		&1.95				&0.89$\pm$0.01		&				&	\\
					&$\nu$			&2132		&2.55				&0.52$\pm$0.01		&870			&0.15\\
					&$\eta$'		&2615		&2.82				&0.87$\pm$0.03		&1400			&0.37\\
					&$\tau$'		&4326		&3.60				&0.98$\pm$0.04		&760			&0.22\\
\hline
\label{TablePar}
\end{tabular}
\end{table}

%\begin{table}
%\caption{Fermi surface parameters obtained from the quantum oscillations for two particular angular orientations of the magnetic field. $F$ is the inverse of the frequency, $k_F$ the Fermi wavevector, $m*$ the effective mass, $\ell(\AA)$ the mean free path, $\tau$ the scattering time and $\mu$ the carrier mobility.}
%\begin{tabular}{c ccccccc}
%
%					&				&F(T)		&$k_F$(nm$^{-1}$)	&m$^*/m_e$				&$\tau$(ps)		&$\mu$(cm$^2$/(V$\cdot$s))	&$l$($\AA$)	\\
%\hline
%8.3$^{\circ}$		&$\beta$		&388		&1.09				& 0.39$\pm$0.03			&0.14			&630		&420	\\
%					&$\gamma$	&1235		&1.94				&0.77$\pm$0.01			&0.51			&1180		&1500	\\
%					&$\eta$		&3023		&3.03				&0.90$\pm$0.03			&0.26			&510		&1020	\\
%					&$\tau$		&4442		&3.67				&0.83$\pm$0.04			&0.73			&1560		&3770	\\
%
%89$^{\circ}$			&$\beta$		&377		&1.04				&0.61$\pm$0.03			&0.26			&750		&520	\\
%					&$\gamma$'	&1323		&2.00				&1.00$\pm$0.01			&0.46			&810		&1070	\\
%					&				&1248		&1.95				&0.89$\pm$0.01			&				&			&		\\
%					&$\nu$			&2132		&2.55				&0.52$\pm$0.01			&0.15			&530		&870	\\
%					&$\eta$'		&2615		&2.82				&0.87$\pm$0.03			&0.37			&760		&1400	\\
%					&$\tau$'		&4326		&3.60				&0.98$\pm$0.04			&				&			&		\\
%\hline
%\label{TablePar}
%\end{tabular}
%\end{table}

If a non-trivial Berry phase $\pi$ is acquired over an orbit, a change in $\gamma$ will occur and the phase of the quantum oscillations will be shifted. In such case, in the linear plot of the Landau index $N_L$ versus $1/B$, the zero $1/B$ intercept of $N_L$ will lie close to 0. In case of a topologically trivial band with no Berry phase accumulation, the zero $1/B$ intercept of $N_L$ will lie inside the interval [3/8,5/8]. In the upper left inset of Fig.\ref{Dingle}(d) we show the oscillation of the $\beta$-orbit at $89^\circ$ (F=377 T). Fig.\ref{Dingle}(d) shows the Landau index $N_L$ plotted as a function of the maximum and minimum position in $1/B$ for this orbit. The zero $1/B$ extrapolation (lower right inset) gives an intercept $N_{L,0} = 0.024\pm0.015$, a value quite close to zero.

%If a Berry phase is acquired over an orbit, this modifies $\gamma$ and the index of the lowest Landau level. Eventually, the lowest Landau level is at $N_L=0$ ($\gamma-\delta$ far from $\frac{1}{2}$ in equation \ref{eq_LK}), instead of at a finite $N_L$ in absence of Berry phase accumulation. For example, in a Dirac dispersion relation with no mass, the lowest Landau level is at $N_L=0$ and there is no zero point energy. We are quite far from the lowest Landau level in all orbits, but we can chose one among the smallest $F$ that also shows large quantum oscillations. In the upper left inset of Fig.\ref{Dingle}(d) we show the oscillation of the $\beta$-orbit at $89^\circ$ (F=377 T). We index the peak and valleys of $\Delta $MR to find $N_L$ and extrapolate the value of $N_L$ to zero $1/B$. We find $N_{L,0} = 0.024\pm0.015$, a value quite close to zero.

\section{Two-band isotropic model including contribution from open orbits}

Let the $\mathbf{z}$-axis be along the direction of the applied magnetic field. Semiclassically, for a free electron gas, the resistivity tensor in the transverse plane ($\mathbf{x}$,$\mathbf{y}$) can be written as: 
\begin{equation}
\hat{\rho} = 
\begin{pmatrix}
\dfrac{1}{\sigma _e} & \dfrac{\eta_e}{\sigma _e}\\[1em]
-\dfrac{\eta_e}{\sigma _e} & \dfrac{1}{\sigma _e}
\label{rho_1band}
\end{pmatrix}
\end{equation}
where $\sigma_e$ is the zero field conductivity and $\eta_e =\mu_eB$ with $\mu_e$ the eletron mobility. In Equation \ref{rho_1band}, $\rho_{xx}=\rho_{yy}$, and it is independent of magnetic field. The corresponding conductivity tensor is the following: 
\begin{equation}
\hat{\sigma}_e = 
\begin{pmatrix}
\dfrac{\sigma _e}{1+\eta_e^2} & -\dfrac{\eta_e\sigma _e}{1+\eta_e^2}\\[1em]
\dfrac{\eta_e\sigma _e}{1+\eta_e^2} & \dfrac{\sigma _e}{1+\eta_e^2}
\end{pmatrix}
\end{equation}
The conductivity tensor for an isotropic single band of hole carriers can be written in the same way, apart from an opposite sign in the off-diagonal terms. When both electrons and holes are present and considered as independent conduction channels, one needs to add them in the conductivity tensor: 
\begin{equation}
\hat{\sigma} = \hat{\sigma}_e +\hat{\sigma}_h =  
\begin{pmatrix}
\dfrac{\sigma _e}{1+\eta_e^2} & -\dfrac{\eta_e\sigma _e}{1+\eta_e^2}\\[1em]
\dfrac{\eta_e\sigma _e}{1+\eta_e^2} & \dfrac{\sigma _e}{1+\eta_e^2}
\end{pmatrix}
+
\begin{pmatrix}
\dfrac{\sigma _h}{1+\eta_h^2} & \dfrac{\eta_h\sigma _h}{1+\eta_h^2}\\[1em]
-\dfrac{\eta_h\sigma _h}{1+\eta_h^2} & \dfrac{\sigma _h}{1+\eta_h^2}
\end{pmatrix}
\label{sigma_2band}
\end{equation}
Inverting $\hat{\sigma}$ leads to the field dependence of the resistivity $\rho_{xx}$ ($x$ being the current direction):
\begin{equation}
\rho_{xx}= \frac{n_e\mu_e+n_h\mu_h+\mu_e\mu_h(n_e\mu_h+n_h\mu_e)B^2}
{e[(n_e\mu_e+n_h\mu_h)^2+(n_e-n_h)^2\mu_e^2\mu_h^2B^2]} 
\label{rhoxx}
\end{equation}
where we have used the Drude formula for conductivity, $\sigma =ne\mu$. Equation \ref{rhoxx} is a result discussed in Ref.\,\onlinecite{Ali2014x}. When the electron and hole carrier numbers come close to compensation, $\rho_{xx}$ reaches a resonance and increases as $B^2$ without saturation. 

In $\gamma$-PtBi$_2$, we can approximate the band structure by a two-band model. In our model, we consider one band which is fundamentally different from the others, due to the topology of its Fermi surface. It has an open Fermi surface formed by sphere-like sheets interconnected with each other, giving a 2D grid in reciprocal space. Electrons can be driven into open, non-circular orbits in the grid. The open orbits extend along the intersection line (along $\mathbf{x}$, the current direction) between the plane perpendicular to the field ($\mathbf{x}$,$\mathbf{y}$), and the [001] plane. Electron dynamics in open orbits is governed only by the usual scattering processes that are present without magnetic field. Thus they lead to an additional field-independent term in $\sigma_{yy}$. With this, we can write:
\begin{equation}
\hat{\sigma} = 
\begin{pmatrix}
\dfrac{\sigma _0}{1+\eta^2} & -\dfrac{d\eta\sigma _0}{1+\eta^2}\\[2em]
\dfrac{d\eta\sigma _0}{1+\eta^2} & \delta\sigma_0 + \dfrac{\sigma _0}{1+\eta^2}
\end{pmatrix}
\label{sigma_oo}
\end{equation}
where $\sigma_0 = \sigma _e+\sigma_h = (n_e+n_h)e\mu$ is the total conductivity at zero field. The two-band model is characterized by $d= (\sigma_e-\sigma_h)/\sigma_0 = (n_e-n_h)/(n_e+n_h)$. The contribution from the open orbits, $\sigma_{o.o.} =\delta\cdot \sigma_0$ is small ($\delta \ll 1$), as only a tiny proportion of the whole Fermi surface should be engaged in forming the open orbits. From Equation \ref{sigma_oo} we deduce the following expression used in the main text, by inverting the conductivity tensor:
\begin{equation}
\rho(B)/\rho_0=\frac{\delta(1+\eta^2)^2+(1+\eta^2)}{\delta (1+\eta^2)+1+ d^2\eta^2}
\label{rho_xx}
\end{equation}

\begin{figure}[h!]
\includegraphics[width =0.9\columnwidth]{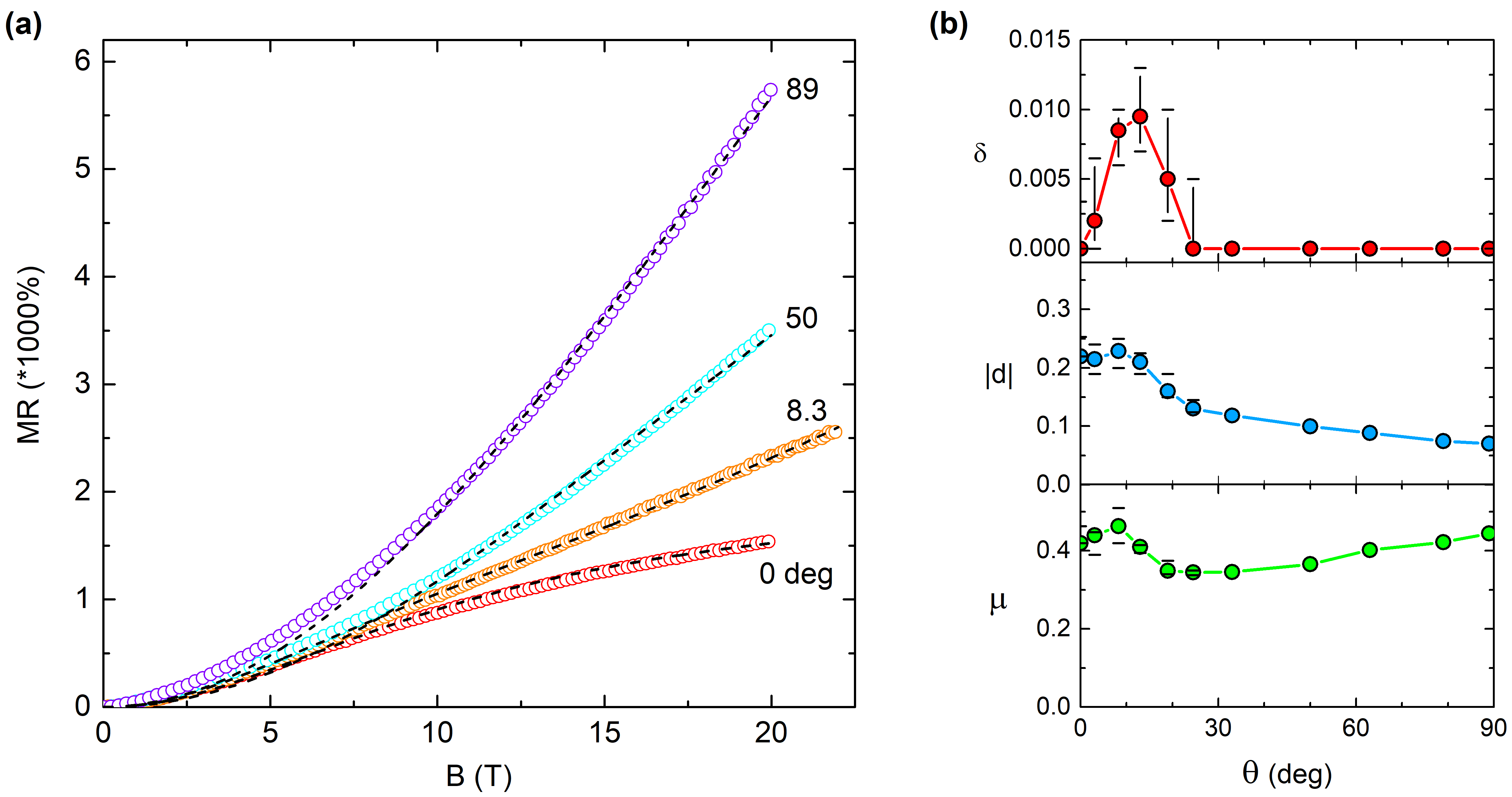}
\caption{
\textbf{(a)} The open circles show the MR data at different field orientations. The dashed lines show the corresponding fits to Equation \ref{rho_xx}. \textbf{(b)} Angular evolution of the used parameters. The contribution from the open orbits $\delta$ appears at $\theta$ close to $8.3^\circ$. The steady increase of the MR with $\theta$ in the whole angular range is mostly associated to a steady decrease of $|d|$ from 0.22 to 0.07 (gradual approaching to an exact electron-hole compensation).}
\label{FigFit}
\end{figure}

Fig.\,\ref{FigFit} (a) shows how Equation \ref{rho_xx} can explain our MR data in the whole angular range. The classical two-band model without open orbit contributions (without $\delta$ terms in Equation \ref{rho_xx}) gives a good account for the MR behavior at most of the angles. In particular, it explains well both the saturating behavior at $\theta = 0^\circ$, and the quasi $B^2$ behavior at $\theta$ close to $90^\circ$. However, at $\theta$ around $8.3^\circ$, it is not possible to reproduce the observed linear or quasi linear MR with the two-band model alone. Notably, a pronounced saturating behavior in the high field limit would always appear in this model. Introducing a contribution from open orbits ($\delta$ terms in Equation \ref{rho_xx}) successfully reproduces the MR. In particular the linear MR at $\theta = 8.3^\circ$ is well explained. Fig.\,\ref{FigFit} (b) shows the angular evolution of the different parameters used in the fits. The steady increase of the MR with $\theta$ in the whole angular range, together with the reversal of the MR curvature, can be associated to a steady decrease of $|d|$ from 0.22 to 0.07 (the eletron and hole number gradually approach compensation). The open orbits contribution to the conductivity $\delta$ equals 0 except at $\theta$ around $8.3^\circ$. The mobility $\mu$ undergoes little change with the angle. 

\section{Angular dependence of the magnetoresistance}

\begin{figure}[h!]
\includegraphics[width = \columnwidth]{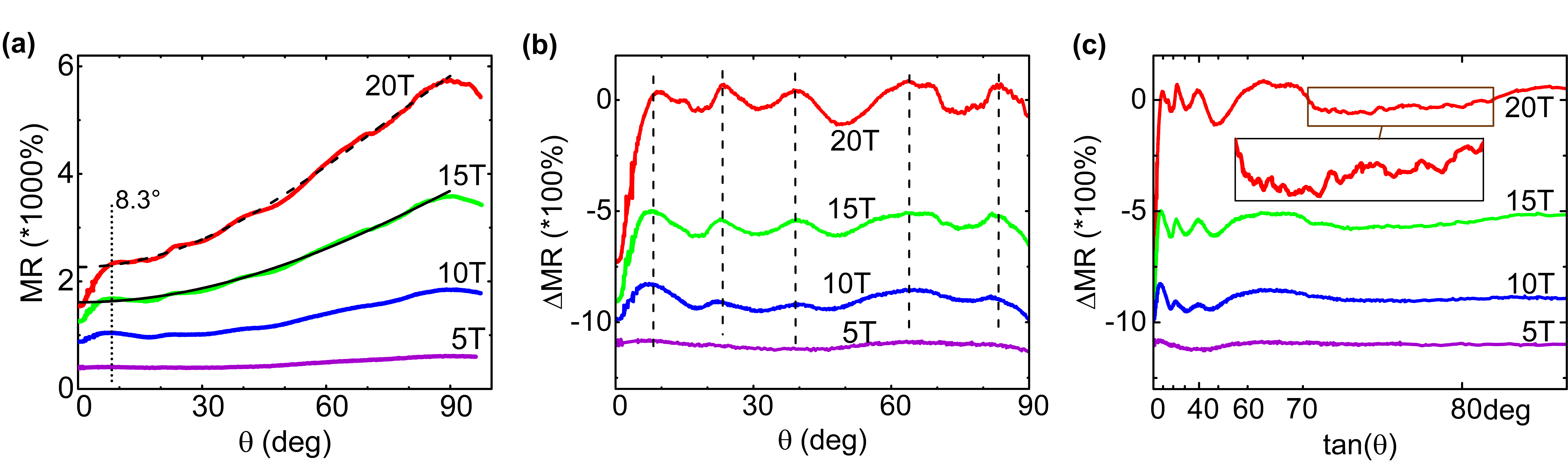}
\caption{\textbf{(a)} Angular dependence of the MR at different magnetic fields, at T=1 K. The vertical dotted line marks the angle where a linear MR is observed: $\theta = 8.3^{\circ}$. The black lines show pair polynomial fits of order 4 to remove the background and get the oscillating part in (b). 
\textbf{(b)} The oscillating part of the angular dependence of the MR at different magnetic fields, at T=1 K (shifted along the vertical axis for clarity). The dashed vertical lines are guides for the eyes, and show that the angles where we observe oscillations do not depend on the magnetic field. 
\textbf{(c)} $\Delta$MR as a function of $\tan (\theta )$, at different magnetic fields. The inset gives a zoom on the signal in the MR due to quantum oscillations, which is strongly magnetic field dependent. }
\label{FigAngDep}
\end{figure}

In Fig.\,\ref{FigAngDep} (a) we show the angular dependence of the MR at constant magnetic fields. We find a strong increase for $\theta$ close to $0^\circ$ which ceases at about $8.3^{\circ}$, the angle at which the MR changes from a saturating to a non-saturating behavior. The dip turns into a MR$(\theta)$ consisting of a smooth background and a small oscillatory component for $\theta>8.3^{\circ}$. We fit the background with a polynomial function of order 4 (dashed lines in Fig.\,\ref{FigAngDep}(a)) and visualize the oscillatory dependence in Fig.\,\ref{FigAngDep}(b). In Fig.\,\ref{FigAngDep}(c) we plot the same data as a function of $\tan (\theta)$. We observe that the magnitude of the oscillations in $\Delta$MR$(\theta)$ increases with the magnetic field. Their position in $\theta$, however, is independent of the magnetic field. Notice that Shubnikov-de Haas quantum oscillations may provide oscillations in the angular dependence of the MR. However, their positions in angle should depend on the field, which is not the case here. Moreover, the amplitude of quantum oscillations in our data is one order of magnitude smaller than the oscillating pattern in the angular dependence of the MR, as we can see in the inset of Fig.\,\ref{FigAngDep}(c). 

Angular dependent MR oscillations, whose position in $\theta$ does not change with the magnetic field, have been obtained in a number of systems\cite{Tafti2015x,Wang2015,Wang2016,Gao2017x,Chen2018,Ren2010,Gabani2003,Kartsovnik2004,Yakovenko2006}. In particular, in layered materials, MR oscillations appear due to two-dimensional Fermi surface sheets with some warping\cite{Kartsovnik2004,Yakovenko2006}. Two-dimensional Fermi surface tubes without warping only have one orbit for the magnetic field applied exactly parallel to the tube. 
%Three-dimensional Fermi surfaces usually show one orbit defining an extremal area, or a multiplicity of these, for all orientations of the magnetic field. 
By contrast, tubes with a small amount of warping show exactly two orbits defining two extremal areas of the Fermi surface for all orientations of the magnetic field. Yamaji showed that the difference between these extremal areas becomes zero at integer values of $\tan(\theta)$\cite{Yamaji1989}. It was then shown that this causes maxima in the angular dependent MR at angles corresponding to integer values of $\tan(\theta)$\cite{Yamaji1989}. Of course, this assumes that warping follows a spherical or cosine-like form along the $\mathbf{c}$-axis. In $\gamma$-PtBi$_2$, quantum oscillations do not show tube-like Fermi surface sheets associated with a 2D nature of a layered system, and there is no sign of them in calculations neither\cite{PhysRevB.94.165201x,Gao2018x}. Instead, they show 3D spherical-like Fermi surfaces sheets with complex geometries. Then the angular-dependent oscillations of the MR may come from a similar effect within a more complex band structure, associated with 3D Fermi surface sheets with a spherical-like geometry which provide orbits with exactly the same area for different angles $\tan (\theta)$. Thus, the origin of the angular dependent oscillations may come from coincidences in size between different orbits on 3D Fermi surface sheets at certain field orientations.

\section{description of the mechanical rotator used for the angular depedent magnetoresistance measurements}

\begin{figure}[h!]
\includegraphics[width =0.6\columnwidth]{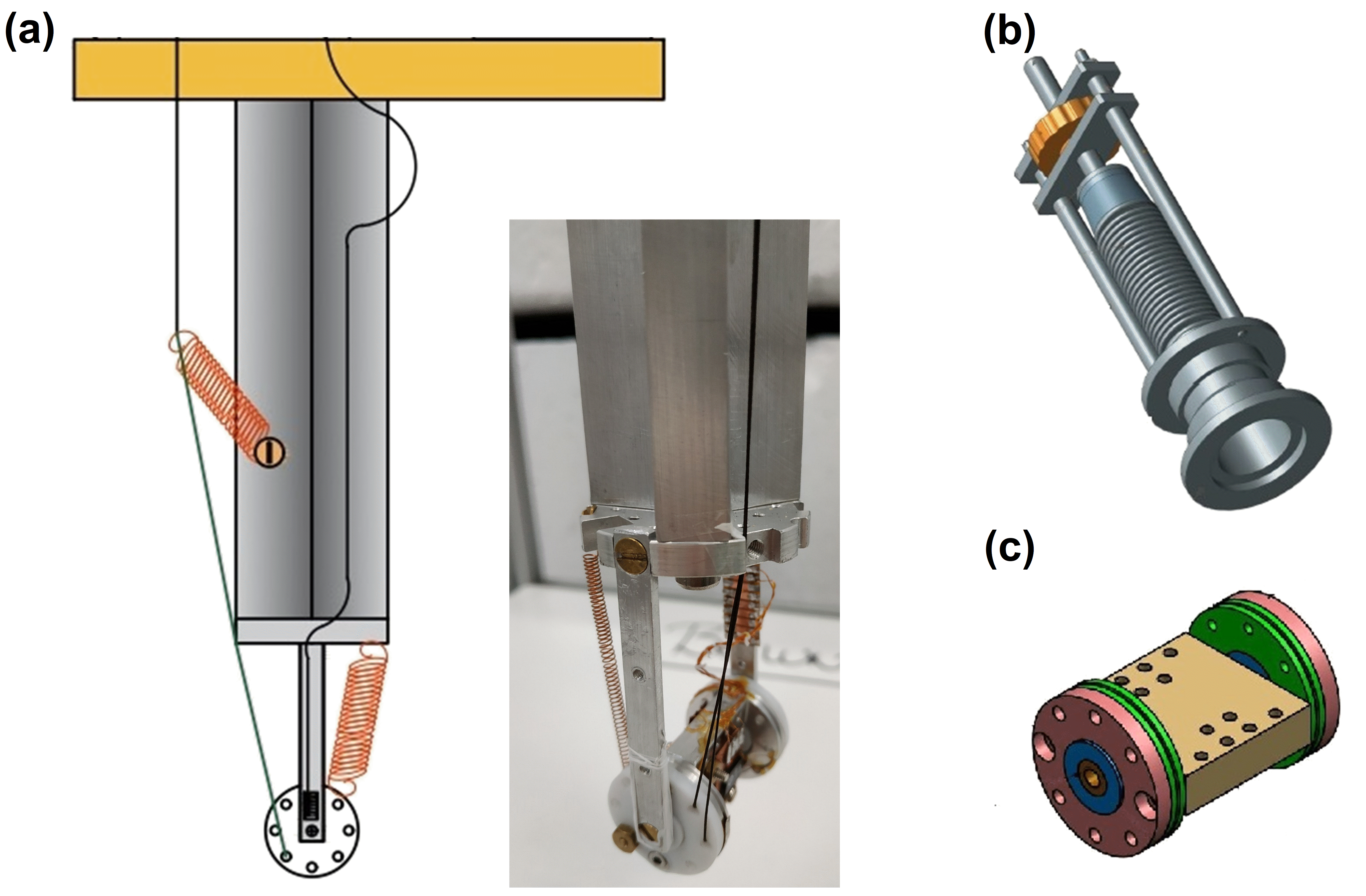}
\caption{
\textbf{(a)} Scheme and photograph of the rotator. \textbf{(b)} Mechanical pulling systems at room temperature. \textbf{(c)} Detail of the rotator. }
\label{Rotator}
\end{figure}

Available designs for rotators consist of gears operated through long tubes going to room temperature. Although the precision achieved with such rotators is very high, friction is often a problem and provides considerable heating at very low temperatures\cite{doi:10.1063/1.1149571}. A solution is to use saphire contacts with very low friction coefficients, although high rotation speeds still produce a sizeable power\cite{doi:10.1063/1.1149571,doi:10.1063/1.4896100}. An additional difficulty is that mechanical design has to consider carefully differential contraction of the whole set-up to avoid clamping at low temperatures. Piezo-driven systems have the advantage of providing compact designs on very small sizes and are commercially available\cite{doi:10.1063/1.5083994,doi:10.1063/1.3502645,doi:10.1063/1.1347982}. However, the application of high frequency signals to the piezoelectrics is nearly unavoidable, which also implies heating. Particularly for high speeds, heating can be a considerable problem, easily reaching the mW range. In addition, piezomotors rely on stick-slip motion, which is difficult to achieve at low temperatures. In another design used in some laboratories, a rotator is operated through Kevlar strings attached to a spring and a shaft that goes to room temperature\cite{Hannahs2010}. This is similar to the wire based nanoscale sample positioning system described in Ref.\onlinecite{doi:10.1063/1.3567008}, which allows for precise cryogenic motion at low temperatures. Here we describe a rotating sample stage devoid of frictional heating. The rotator is operated by pulling on a string that goes to the top of the experiment (Fig.\,\ref{Rotator}(a)), through a bellow at ambient temperature (Fig.\,\ref{Rotator}(b)). The string is thermalized at each stage of the refrigerator and is maintained straight by using a spring connected to the support mechanism. At the level of the rotator, there is another spring connected to the support. It pulls the rotator back to the initial position when the tension in the string is released. The rotator is mounted with a copper cylinder as the rotation axis (Fig.\,\ref{Rotator}(c)). The copper cylinder is firmly anchored at the end of the support. The pieces surrounding the cylinder are of teflon to allow for smooth rotation without generating heat. We have built rotators in Macor, Aluminum and plastic, details and drawings are provided in Ref.\onlinecite{OSF}.

To make the resistivity measurements, we use a four probe configuration and a conventional lock-in amplifier to measure the voltage. To apply the current, we use a Howland pump circuit \cite{OSF}. We apply a voltage to the sample through a resistor $R$. The voltage is sensed at the exit of $R$ by an operational amplifier and maintained constant, independently of the impedance of the sample, by a feedback driven through another operational amplifier. The current range can be modified by modifying $R_x$ with a switch. The current range is given by the input voltage and the resistor $R$\cite{HowlandTI,doi:10.1063/1.4878255,HowlandHammon}. The impedance of the source increases as the inverse of the precision in the resistors used. Thus, the sensitivity to modifications in the current is considerably improved with respect to the use of a simple resistor in series and a voltage source. For instance, using 0.01\% resistors implies source impedances multiplied by two orders of magnitude. The bandwidth of the circuit is limited by the operational amplifiers and a capacitor inserted parallel to $R$ and is in any case above 10kHz.

%\bibliographystyle{apsrev4-1-titles}
%\bibliography{bib}


\end{document}